\documentclass{article}
\usepackage{amsmath}[1996/11/01]
\usepackage{amssymb,amsthm,amsxtra}
\usepackage{epsfig}

\newlength{\mytopmargin}
\newlength{\myleftmargin}
\def\zz{\rlx\hbox{\small \sf Z\kern-.4em Z}}

\newtheorem{lemma}{Lemma}[section]

\newtheorem{prop}[lemma]{Proposition}
\newtheorem{cor}[lemma]{Corollary}

\begin{document}

\vspace{1cm}
\noindent
\begin{center}{   \large \bf Particles in a magnetic field and plasma analogies: \\
doubly periodic boundary conditions}
\end{center}
\vspace{5mm}

\noindent
\begin{center}
 P.J.~Forrester\\

\it Department of Mathematics and Statistics, \\
University of Melbourne, Victoria
3010, Australia
\end{center}
\vspace{.5cm}
\begin{quote}
The $N$-particle free fermion state for quantum particles in the
plane subject to a perpendicular magnetic field, and with doubly periodic
boundary conditions, is written in a product form. The absolute value of
this is used to formulate an exactly solvable one-component plasma model, and
further motivates the formulation of an exactly solvable two-species Coulomb
gas. The large $N$ expansion of the free energy of both these models exhibits
the same $O(1)$ term. On the basis of a relationship to the Gaussian free
field, this term is predicted to be universal for conductive Coulomb systems
in doubly periodic boundary conditions.
\end{quote}

\vspace{.5cm}
\noindent
\section{Introduction}
\setcounter{equation}{0}
The two-dimensional one-component plasma is a model of two-dimensional point charges
in equilibrium. The point charges all have the same sign and
magnitude, $+q$ say, and the system is neutralized by a uniform smeared out negative
background, of total charge $-qN$. The total potential energy then consists of a
particle-particle contribution, a particle-background contribution, and a
background-background contribution. Using the fact the solution of the
two-dimensional Poisson equation in free boundary conditions is given by
$- \log |\vec{r}_1 - \vec{r}_2|$, for a system of $N$ point charges with
coordinates $\vec{r}_1,\dots, \vec{r}_N$ confined to a neutralizing disk of
unit charge density centred about the origin, the corresponding Boltzmann factor
is readily calculated to be proportional to (see e.g.~\cite{Fo98a})
\begin{equation}\label{1.2dOCP}
e^{- \pi \Gamma  \sum_{j=1}^N |\vec{r}_j|^2/2}
\prod_{1 \le j <k \le N} |\vec{r}_k - \vec{r}_j|^\Gamma, \qquad \Gamma := q^2/k_B T.
\end{equation}
Introducing the complex coordinates $z_j = x_j + i y_j$ for
$\vec{r}_j = (x_j, y_j)$, and recalling the Vandermonde determinant formula
\begin{equation}
\det [ z_j^{k-1} ]_{j,k=1,\dots,N} = \prod_{1 \le j < k \le N}
(z_k - z_j),
\end{equation}
one sees that (\ref{1.2dOCP}) in the case $\Gamma = 2$ can be written
\begin{equation}\label{1.1a}
\Big | \det [ \psi_k(z_j) ]_{j,k=1,\dots,N} \Big |^2
\end{equation}
where
\begin{equation}\label{2}
\psi_k(z) := e^{- \pi |z|^2/2} z^{k-1}.
\end{equation}
Up to normalization, this is the absolute value squared of a free Fermi
system of $N$ particles for which the single particle wave functions in state
$k$ are proportional to (\ref{2}).

The wave functions (\ref{2}) are realized as degenerate eigenfunctions in the
lowest energy level (Landau level) for a single quantum particle confined to the
$xy$-plane and subject to a perpendicular magnetic field $\vec{B} =
B \hat{z}$, $B > 0$. To revise this point
(see e.g.~\cite{CDL77}), one notes that the corresponding
Hamiltonian is
\begin{equation}\label{2.1}
H = {1 \over 2m} ( - i \hbar \nabla + {e \over c} \vec{A} )^2
\end{equation}
where the vector potential $\vec{A}$ must satisfy
$$
\nabla \times \vec{A} = B \hat{z}.
$$
The mechanism for the degeneracies in (\ref{2.1}) is the fact that $H$ commutes
with the quantum analogue of the square of the classical orbit centre,
$X^2 + Y^2$, where
$$
X=x-{{l^2}\over\hbar}\Pi_y,\quad Y=y+{{l^2}\over\hbar}\Pi_x
$$
with $l:=\sqrt{\hbar c/eB}$ the magnetic length and
$$
\Pi_x=-i\hbar{\partial\over{\partial x}}+{e\over c}\,A_x,\quad
\Pi_y=-i\hbar{\partial\over{\partial y}}+{e\over c}\, A_y
$$
generalised momenta. A standard calculation shows that the energy levels of
(\ref{2.1}) are  $E_n=(n+{1\over 2})\hbar w_c$,
$w_c:=eB/mc$, for $n=0,1,2,\dots$ while the eigenvalues of $X^2 + Y^2$ are
equal to $(2m+1)^2$, $m=0,1,2,\dots$. Moreover, for the ground state energy
$E_0 = {1 \over 2} \hbar \omega_c$ the eigenfunctions of $X^2 + Y^2$ are
\begin{equation}\label{3.1}
\psi_m(\vec{r})={{\bar{z}}^m e^{-(x^2+y^2)/4l^2}
\over{{(2\pi l^22^ml^{2m}m!)}^{1/2}}}.
\end{equation}

The states (\ref{3.1}) are mutually orthogonal, and have the interpretation that they
have definite values for the distance from the origin to the centre of their
cyclotron orbit (implied by the eigenvalue of $X^2 + Y^2$) which increases with
$m$. The most dense $N$-particle state $\psi$, in which the particles are fermions
but otherwise non-interacting, is therefore obtained by constructing a Slater
determinant from the states $\psi_0(\vec{r}), \dots, \psi_{N-1}(\vec{r})$, and thus
up to normalization its absolute value squared corresponds precisely to
to that specified by (\ref{1.1a}) and (\ref{2}).

The aim of this paper is to develop the plasma analogy for the same quantum problem
as above, but now with doubly periodic boundary conditions. This suggests a solvable
version of the doubly periodic two-dimensional one-component plasma. For this
particular model (which involves an $N$-body potential), the leading finite size
correction to the bulk free energy can be computed exactly.
Having a solvable doubly periodic plasma system identified by the quantum problem,
we consider the corresponding version of the two-dimensional Coulomb gas
(mobile positive and negative point charges) at the coupling $\Gamma = 2$. The
leading finite size correction to the pressure in the expansion of the grand
potential can be computed exactly. It is found to coincide with the correction
term obtained for the one-component plasma. We end with a discussion of the
universality of this term for two-dimensional Coulomb systems in doubly
periodic boundary conditions and in 
their conductive phase.

\section{The one-component plasma in doubly periodic boundary conditions}
\setcounter{equation}{0}
The two-dimensional Poisson equation reads
\begin{equation}\label{5.0}
{\partial^2 \tilde{\Phi} \over \partial x^2} +
{\partial^2 \tilde{\Phi} \over \partial y^2} =
- 2 \pi \delta(x - x') \delta (y - y').
\end{equation}
Making use of complex coordinates,
in free boundary conditions it has the solution
\begin{equation}\label{5.1}
\tilde{\Phi}(z,z') = - \log |z - z'|
\end{equation}
We seek the solution satisfying doubly periodic boundary conditions
\begin{eqnarray}\label{5a}
 \tilde{\Phi}((x+L,y),(x',y')) & = & \tilde{\Phi}((x,y),(x',y')) \\
\tilde{\Phi}((x,y+W),(x',y')) & = & \tilde{\Phi}((x,y),(x',y')).
\label{5b}
\end{eqnarray}

For this task, following \cite{Fo90d}, consider the Jacobi theta function
\begin{eqnarray}\label{11.th1}
\theta_1(z;q) & = & - i \sum_{n = - \infty}^\infty (-1)^n q^{(n - 1/2)^2}
e^{2 i (n - 1/2) z} \nonumber \\
& = & 2q^{1/4}\sin
z\prod_{n=1}^\infty\left(1-q^{2n}e^{2iz}\right)\left(1-q^{2n}e^{-2iz}\right)
\left(1-q^{2n}\right).
\end{eqnarray}
The fact that $\theta_1$ is an entire function which vanishes if and only if
$z = \pi m + \pi \tau n$, $m,n \in \mathbf Z$,  $q = e^{ i \pi  \tau}$ (Im$(\tau) >0$),
and that $\theta_1(z;q) \sim z \theta_1'(0;q)$ as $z \to 0$, tells us that
\begin{equation}\label{6.1}
\tilde{\Phi}(z,z'):=-\log\left({{L|\theta_1(\pi(z-z')/L;q)|}\over
{\pi\theta_1'(0;q)}}\right), \quad q:=e^{-\pi W/L}
\end{equation}
satisfies the Poisson's equation (\ref{5.0}) for $0\leq x,x'<L$, $0\leq
y,y'<W$, with the further specification that (\ref{5.1}) holds as 
$|z - z'| \to 0$. Moreover, since
\begin{equation}\label{11.t1p}
\theta_1(z + \pi;q) = - \theta_1(z;q) \quad {\rm and} \quad
\theta_1(z + \pi \epsilon;q) = - q^{-1} e^{-2 i z} \theta_1(z;q)
\end{equation}
we see that (\ref{6.1}) obeys (\ref{5a}), while (\ref{5b}) must be modified to read
\begin{equation}
\tilde{\Phi}((x,y+W),(x',y'))  =  -{\pi \over L}(2y+W) +
\tilde{\Phi}((x,y),(x',y')).
\end{equation}
This latter point is of no surprise, as it is not possible to solve the Poisson
equation in doubly periodic boundary conditions unless it is made charge
neutral and thus modified to read
\begin{equation}\label{6.2}
{\partial^2 \tilde{\Phi} \over \partial x^2} +
{\partial^2 \tilde{\Phi} \over \partial y^2} =
- 2 \pi \delta(x - x') \delta (y - y') + {2 \pi \over LW}.
\end{equation}
From the above working, the doubly periodic solution of 
(\ref{6.2}) is seen to be
\begin{equation}\label{6.3}
\Phi(z,z') = {\pi y^2 \over LW} + \tilde{\Phi}(z,z').
\end{equation}

From the viewpoint of the magnetic analogy, it is of interest to construct a
2dOCP from the quasi-periodic potential (\ref{6.1}) rather than the fully
periodic potential (\ref{6.3}). Consider then a charged system of $N$ mobile
particles, charge $+q$, confined to the rectangle $0 < x < L$, $0 < y < W$,
interacting via the pair potential (\ref{6.1}). Also present is a smeared out
uniform background of total charge density $-N/LW$. The corresponding
particle-background potential is given by
\begin{equation}\label{7.1}
U_2 := q^2 \sum_{j=1}^N V(z_j) \qquad {\rm where} \qquad
V(z) = \int_0^L dx' \int_0^W dy' \, \Phi(z,z'),
\end{equation}
while for the background- background interaction we have
$$
U_3 := - {q^2 \over 2} {N \over LW} \int_0^L dx \int_0^W dy \, V(z).
$$
For the integral in (\ref{7.1}), according to (\ref{6.1}) we must evaluate
$$
I(y')=\int_0^Ldx\int_0^Wdy\log\left|\theta_1\left(\pi(x-x')/L+\pi
i(y-y')/L;e^{-\pi W/L}\right)\right|.
$$
Use of the product form in (\ref{6.1}) shows
$$
I(y')={{LW}\over3}\log\Big ({1\over
2}\theta_1'(0;q)\Big )+\pi{\Big (y'-{W\over2}\Big )}^2+{\pi{W^2}\over{12}}.
$$
With $U_1$ denoting the particle-particle interaction, the Boltzmann factor is thus
\begin{eqnarray}\label{8.1}
&& e^{-\beta (U_1 + U_2 + U_3)} 
= {\left({{\pi\theta_1'(0;q)}\over L}\right)}^{N\Gamma/2}
\nonumber \\
&&
\times e^{-(\Gamma
N^2/6)\log(\theta_1'(0;q)/2)}e^{-\pi\rho\Gamma\sum_{j=1}^N
{(y_j-W/2)}^2}\prod_{1\leq
j<k\leq N}{|\theta_1(\pi(z_k-z_j)/L;q)|}^\Gamma
\end{eqnarray}

\section{Magnetic analogy}
\subsection{The $N$-particle wave function}
\setcounter{equation}{0}
We seek to construct the analogue of the states (\ref{3.1}), and then to compare
(\ref{1.1a}) with the Boltzmann factor (\ref{8.1}). For the first of these tasks,
we follow \cite{HR85}. The double periodicity associated with (\ref{6.3}) has as
its fundamental domain a rectangle. In discussing the Hamiltonian (\ref{2.1})
in doubly periodic boundary conditions it is more natural to take the
fundamental domain as a parallelogram with corners at $(0,0), (0,L),
(L+W_1,W_2), (W_1,W_2)$. But independent of this detail, it is not possible to
construct a vector potential which is doubly periodic and satisfies 
$\nabla \times \vec{A} = B \hat{\mathbf z}$. Instead two vector potentials
$A^W$ and $A^L$,
related by the gauge transformation $A^W = A^L + \nabla f$, are defined.
The vector potential $A^L$ is periodic under $x \mapsto x + L$,
while $A^W$ is periodic under $x \mapsto x + W_1$, $y \mapsto y + W_2$.
From these potentials corresponding ground state solutions
$\psi^L$ and $\psi^W$ can be constructed which satisfy
\begin{equation}\label{11.59}
\psi^L(x,y)=\psi^L(x+L,y)\ {\rm{and}}\ \psi^W(x,y)=\psi^W(x+W_1,y+W_2),
\end{equation}
and furthermore are related by
\begin{equation}\label{11.61}
\psi^W(x,y)=\psi^L(x,y)e^{-(ie/\hbar c)f(x,y)}.
\end{equation}
This latter fact follows from $A^W$ and $A^L$ being related by a gauge transformation.
Note that the absolute values of $\psi^W$ and $\psi^L$ are equal and have the
periodicity of the parallelogram.

For the vector potentials we take
$$A^L=-By\hat{x},\ A^W={B\over
2}\left(\left({{W_2}\over{W_1}}x-y\right)\hat{x}+\left(x-{{W_1}\over{W_2}}y
\right)\hat{y}\right),$$
which have the periodicity
required by (\ref{11.59}) and are related by a gauge
transformation with the scalar function given explicitly by
\begin{equation}\label{11.60}
f(x,y)={B\over
2}\left({{W_2}\over{2W_1}}x^2+xy-{{W_1}\over{2W_2}}y^2\right).
\end{equation}
Substituting the property (\ref{11.61}) in the periodicity equation for
$\psi^W$ gives
\begin{equation}\label{11.60a}
\psi^L(x,y) = \psi^L(x+W_1,y+W_2) e^{-i W_2 (2x+W_1)/2l^2}.
\end{equation}
But $\psi^L$ is periodic in $x$ of period $L$ so for solutions with the
properties (\ref{11.59}) the magnetic field must be such that
\begin{equation}\label{11.61b}
W_2={{2\pi l^2 N}\over L}
\end{equation}
for some $N=1,2,\dots$.  Since the parallelogram has area $LW_2$, this
condition says that the total magnetic flux $BLW_2$ is an integer
multiple of the flux quanta $\Phi_0=hc/e$.

An easy to establish general property of the Hamiltonian  (\ref{2.1}) is that
the lowest energy state periodic in $x$ can be written in the form
$f(e^{2 \pi i \bar{z}/L}) e^{-y^2/2 l^2}$ where $f(u)$ is a Laurent series in $u$,
and so we can write
\begin{equation}\label{11.63}
\Psi^L(x,y)=e^{-y^2/2l^2}\sum_{n=-\infty}^\infty a_n
e^{-2\pi i n \bar{z}/L}.
\end{equation}
Substituting (\ref{11.63}) in (\ref{11.60a})
 allows a simple recurrence for $a_n$ to be obtained. There are $N$ independent
solutions which when substituted back in (\ref{11.63}) gives
\begin{equation}\label{11.64}
\psi_m^L(x,y)={{e^{-y^2/2l^2}}\over{\sqrt{Ll\pi^{1/2}}}}q^{m^2/N}e^{-2\pi
im \bar{z}/L}\theta_3\left(\pi(\tau m-N\bar{z}/L);q^N\right),
\end{equation}
where $q := e^{\pi i \tau}$ with
$\tau=(-W_1+iW_2)/L$, $m=0,1,\dots,N-1$ and
\begin{equation}\label{11.th3}
\theta_3(u;q) := \sum_{n = - \infty}^\infty q^{n^2} e^{2 i u n}.
\end{equation}

From the $N$ single particle states (\ref{11.64}), the $N$-particle free Fermi
state in (\ref{1.1a}) is formed. To relate it to the Boltzmann factor
(\ref{8.1}) requires the following generalizations of the Vandermonde
identity \cite{Fo90a}.

\begin{prop}\label{pot}
With $(q^2;q^2)_\infty = \prod_{j=1}^\infty (1 - q^{2j})$,
let 
$$
f_N(q)  := N^{N/2} q^{-(N - 1)(N -2)/24} (q^2;q^2)^{-(N-1)(N-2)/2}_\infty.
$$
For $N$ odd
\begin{eqnarray}\label{11.odd}
\lefteqn{
\det \Big [ \theta_3 \Big ( \pi(x_j + \alpha - l/N); q^{1/N} \Big )
\Big ]_{j,l=1,\dots,N}} \nonumber \\ && =
\theta_3 \Big ( \pi \sum_{j=1}^N(x_j + \alpha);q \Big ) f_N(q)
\prod_{1 \le j < k \le N} \theta_1 \Big ( \pi (x_k - x_j);q \Big )
\end{eqnarray}
while for $N$ even
\begin{eqnarray}\label{11.even}
\lefteqn{
\det \Big [ \theta_1 \Big ( \pi(x_j + \alpha - l/N); q^{1/N} \Big )
\Big ]_{j,l=1,\dots,N} } \nonumber \\ && =
\theta_4 \Big ( \pi \sum_{j=1}^N(x_j + \alpha);q \Big ) f_N(q)
\prod_{1 \le j < k \le N} \theta_1 \Big ( \pi (x_k - x_j);q \Big ).
\end{eqnarray}
In (\ref{11.odd}), (\ref{11.even}), $\theta_1$ is specified by (\ref{11.th1}),
$\theta_3$ by (\ref{11.th3}), while
\begin{equation}\label{11.th4}
\theta_4(u;q) = \sum_{n=-\infty}^\infty (-1)^n q^{n^2} e^{2 \pi i n}.
\end{equation}
\end{prop}

The determinants in Proposition \ref{pot} can be transformed into the form required by
the free Fermi state in (\ref{1.1a}), $\psi^{\rm DP}$ say,
corresponding to (\ref{11.64}). This is done by multiplying both sides of
(\ref{11.odd}) with $\alpha = 0$ by
$$
\det [ e^{2 \pi i l k /N} ]_{l=1,\dots,N \atop k=0,\dots,N-1} =
N^{N/2} i^{(N-1)( 3N/2 + 1)}
$$
and multiplying both sides of (\ref{11.even}) with $\alpha = - \pi \tau /2$
by
$$
\det [ e^{2 \pi i l (k + 1/2) /N} ]_{l=1,\dots,N \atop k=0,\dots,N-1} =
N^{N/2} i^{N + 1} i^{(N-1)( 3N/2 + 1)}.
$$
As a consequence the $N$-particle state $\psi^{\rm DP}$, defined as a 
determinant of theta functions
\begin{eqnarray}\label{11.65}
\psi^{\rm DP}((x_1,y_1),\dots,(x_N,y_N)):={1\over{\sqrt{N!}{(Ll\sqrt{\pi})}^{N/2}}}
\exp\Big (-\sum_{j=1}^Ny_j^2/2l^2\Big )q^{\sum_{m=0}^{N-1}m^2/N} \nonumber \\
\hspace*{1.5cm} \times{\rm{det}}\left[e^{-2\pi
i(k-1)\bar{z}_j/L}\theta_3\left(\pi(\tau(k-1)-N\bar{z}_j/L);q^N\right)
\right]_{j,k=1,\dots,N},
\end{eqnarray}
can be written in the factorized form
\begin{eqnarray}\label{11.66}
\psi^{\rm DP}((x_1,y_1),\dots,(x_N,y_N))={{i^{(N-1)(3N/2+1)}f_N(q)}\over{\sqrt{N!}
{(LNl\sqrt{\pi})}^{N/2}}}\exp\Big (-\sum_{j=1}^Ny_j^2/2l^2\Big ) \nonumber\\
\times\theta_{s}\Big (-\pi\sum_{j=1}^N\bar{z}_j/L;q\Big )\prod_{1\leq
j<k\leq N}\theta_1\left(-\pi(\bar{z}_k-\bar{z}_j)/L;q\right).
\end{eqnarray}
where $s=3$ for $N$ odd and $s=1$ for
$N$ even.

We see that $|\psi^{\rm DP}|^2$ with $W_1 = 0$, $W_2 = W$, $1/l^2 = 2 \pi$ is
closely related to the Boltzmann factor (\ref{8.1}) with $\Gamma = 2$. In
fact the two expressions are proportional except that in (\ref{8.1}) there is
a factor of $e^{2 \pi W \sum_{j=1}^N y_j}$, while in $|\psi^{\rm DP}|^2$ this
factor is replaced by $|\theta_s(-\pi \sum_{j=1}^N \bar{z}_j/L;q)|^2$.
Equivalently, since
\begin{equation}
e^{-2 \pi \rho \sum_{j=1}^N y_j^2} \Big |
\theta_s \Big ( -\pi\sum_{j=1}^N\bar{z}_j/L;q\Big )\Big |^2
= e^{-2 \pi \rho \sum_{j=1}^N (y_j - W/2)^2} \Big |
\theta_1 \Big ( \pi\sum_{j=1}^N (\bar{z}_j - (L - iW)/2)/L;q\Big )\Big |^2
\end{equation}
we see that $|\psi^{\rm DP}|^2$ differs from (\ref{8.1}) with $\Gamma = 2$ by a
constant factor times the many body term
\begin{equation}\label{11.122}
\Big | \theta_1 \Big ( \pi\sum_{j=1}^N (\bar{z}_j - (L - iW)/2)/L;q\Big )
\Big |^2.
\end{equation}

\subsection{A doubly periodic plasma with $N$-body potential}
Being a free Fermi state, the many body quantum system corresponding to
(\ref{11.65}) is exactly solvable in the sense that its $l$-point
density matrix can be expressed explicitly as an $l \times l$
determinant. The diagonal term of this density matrix gives the
$l$-point ground state  correlation function. This correlation function
is identical to that for the corresponding classical state with Boltzmann
factor proportional to $|\psi^{\rm DP}|^2$. It follows that if the
plasma system with Boltmann factor (\ref{8.1}) at $\Gamma = 2$ is
augmented by multiplication by (\ref{11.122}), a solvable model is obtained. 
Moreover, what was a quasi doubly periodic system now becomes fully
doubly periodic.

Rather than consider the correlation functions, let us consider the free
energy of this system.
The state $\psi^{\rm DP}$ is normalized so that $|\psi^{\rm DP}|^2$
integrated over $0 \le x_j \le L$, $0 \le y_j \le W$ $(j=1,\dots,N)$ gives unity.
It thus follows from (\ref{11.66}) that
\begin{eqnarray}\label{11.123}
\int_0^Ldx_1 \cdots \int_0^Ldx_N \int_0^Wdy_1 \cdots \int_0^Wdy_N \,
 e^{-2\pi\rho \sum_{l=1}^N (y_l - W/2)^2}
{\Big |\theta_{1}\Big (\pi\sum_{j=1}^N(\bar{z}_j-(L - iW)/2)/L;q\Big )\Big |}^2
\nonumber \\ \times \prod_{1\leq
j<k\leq N}
{\left|\theta_1\left(\pi(z_k-z_j)/L;q\right)\right|}^2
=N!{\left(LN{(2\rho)}^{-1/2}\right)}^N{\left(f_N(q)\right)}^{-2}.
\end{eqnarray}
Hence for the partition function of the solvable plasma system we have
\begin{eqnarray*}
Z_N & := & {1 \over N!} \int_0^L dx_1 \cdots \int_0^L dx_N
\int_0^W dy_1 \cdots \int_0^W dy_N \,
\Big | \theta_1 \Big ( \pi\sum_{j=1}^N (\bar{z}_j - (L - iW)/2)/L;q\Big )
\Big |^2 \\
& & \times e^{- \beta (U_1 + U_2 + U_3) } \nonumber \\
& = &
\Big ( {\pi \theta_1'(0;q) \over L} \Big )^N
e^{-(N^2/3) \log (\theta_1'(0;q)/2)} (LN(2\rho)^{-1/2})^N
\Big ( f_N(q) \Big )^{-2} \\ &= &
\pi^N (2 \rho )^{-N/2} q^{1/6} \prod_{k=1}^\infty (1 - q^{2k})^2, \quad
q := e^{- \pi L / W},
\end{eqnarray*}
and consequently the total free energy is given by
\begin{equation}\label{11.fr2p}
\beta F := - \log Z_N =  {N \over 2} \log \rho / 2 \pi^2
+ 2 \log \Big ( q^{1/12} \prod_{k=1}^\infty (1 - q^{2k} ) \Big ).
\end{equation}
One notes that the corresponding free energy per unit volume 
in the thermodynamic limit
is precisely that known
for the plasma system at $\Gamma = 2$ in a disk \cite{AJ81}, and also there is no
surface free energy in keeping with the system being doubly periodic.

\section{The two-dimensional Coulomb gas}
\setcounter{equation}{0}
At the coupling $\Gamma = 2$ the two-dimensional Coulomb gas of equal numbers
of positive and negative point charges in the grand canonical ensemble has been
solved exactly in various boundary conditions 
\cite{CJ89,Fo91d,Fo92f,JT96,JS01}. However these solvability
properties do not carry through to the case of doubly periodic conditions with
pair potential (\ref{6.3}) \cite{Fo90d}. Instead, the experience gained from the study
of the one-component plasma above suggests that we consider a model with the
quasi doubly periodic potential (\ref{6.1}), and extend it to a doubly periodic
model by multiplying by a term analogous to (\ref{11.122}). The latter is chosen to
be
$$
\Big | \theta_4\Big ( {\pi \over L} \sum_{j=1}^N (w_j - z_j) ; q \Big )
\Big |^2
$$
where $w_j$ and $z_j$ denote the complex coordinates of the positive and
negative charges respectively. The total Boltzmann factor at $\Gamma = 2$,
$W_{N2}$ say, is thus given by
\begin{eqnarray}\label{13.mb5}
W_{N2} = \Big ( {\pi \over L} \theta_1'(0;q) \Big )^{2N}
\Big | \theta_4\Big ( {\pi \over L} \sum_{j=1}^N (w_j - z_j) ; q \Big )
\Big |^2
\Big | F(w_1,\dots,w_N; z_1,\dots, z_N;q) \Big |^2, \nonumber \\
 F(w_1,\dots,w_N; z_1,\dots, z_N;q) :=
(-1)^{N(N-1)/2} {\prod_{1 \le j < k \le N}
\theta_1(w_k - w_j;q) \theta_1(z_k - z_j;q) \over
\prod_{j,k=1}^N \theta_1(w_j - z_k;q)}.
\end{eqnarray}
We seek to evaluate the grand partition function
$$
\Xi_2(\zeta) = \sum_{N=0}^\infty \zeta^{2N} {1 \over (N!)^2}
\int_0^L dx_1 \cdots \int_0^L dx_N \int_0^W dy_1 \cdots \int_0^W dy_N  \,
W_{N2}.
$$
For this we require a classical determinant identity due to Frobenius \cite{Fr82}
(see also \cite{Ao94a}).

\begin{prop}
With $F$ specified in (\ref{13.mb5}),
\begin{eqnarray}\label{13.mm5}
\theta_4(\sum_{j=1}^N (w_j - z_j) - \alpha;q)
 F(w_1,\dots,w_N; z_1,\dots, z_N;q)  \nonumber \\
= \theta_4(\alpha;q) \det \Big [
{\theta_4(w_j - z_k - \alpha;q) \over
\theta_4(\alpha;q) \theta_1(w_j - z_k;q)} \Big ]_{j,k=1,\dots,N}.
\end{eqnarray}
\end{prop}

Using (\ref{13.mm5}) with $\alpha = 0$ shows we can write
\begin{equation}\label{st}
W_{N2} =  (\theta_4(0;q))^2
\det \left [
\begin{array}{cc}  0_N & [K(w_j - z_k) ]_{j,k=1,\dots,N} \\{}
[K(\bar{z}_j - \bar{w}_k) ]_{j,k=1,\dots,N} &  0_N
\end{array} \right ]
\end{equation}
where
$$
K(w-z) := 
{\pi \theta_1'(0;q) \over L \theta_4(0;q)} 
{ \theta_4(w - z;q) \over \theta_1(w -z;q)}.
$$
The structure (\ref{st}) is familiar in studies of the two-dimensional Coulomb
gas at $\Gamma = 2$ \cite{CJ87}. It allows the Fredholm theory of integral operators to be
applied \cite{WW65} which tells us that
\begin{equation}\label{13.xfs}
\Xi_2(\zeta) =  (\theta_4(0;q))^2 \det(1 + \zeta \tilde{K}) = 
(\theta_4(0;q))^2
\prod_{\alpha}(1 + \zeta
\lambda_\alpha)
\end{equation}
where $ \tilde{K}$ is the integral operator such that the corresponding
eigenvalues $\lambda$ and
eigenfunctions $\psi_1(x,y)$ and $\psi_2(x,y)$ are specified by the
coupled equations
\begin{eqnarray}\label{13.epp}
\int_0^Ldx_2 \int_0^Wdy_2 \,
\psi_2(x_2,y_2) K(z_1 - z_2) & = &
\lambda \psi_1(x_1,y_1) \nonumber \\
\int_0^Ldx_2 \int_0^Wdy_2 \,
\psi_1(x_2,y_2) K(\bar{z}_1 - \bar{z}_2) 
 & = &
\lambda \psi_2(x_1,y_1).
\end{eqnarray}

As an aside, we note that because (\ref{6.1}) satisfies the Poisson equation
(\ref{5.0}), and further for $0 \le x_1, x_2 \le L$,  $0 \le y_1, y_2 \le W$,
$K(z_1-z_2)$ is analytic except at $z_1 = z_2$, we have
$$
{\partial \over \partial \bar{z} } {\theta_4(z - z') \over 
\theta_1(z - z') } = \pi \delta (x - x') \delta (y - y').
$$
Hence applying $\partial / \partial \bar{z}_1$ to the first equation in 
(\ref{13.epp}), and $\partial / \partial z_1$ to the second equation, reduces the
coupled differential equations
\begin{eqnarray*}
{\pi^2 \theta_1'(0;q) \over L \theta_4(0;q) } \psi_2(x,y) & = & \lambda
{\partial \over \partial \bar{z} } \psi_1(x,y) \\
{\pi^2 \theta_1'(0;q) \over L \theta_4(0;q) } \psi_1(x,y) & = & \lambda
{\partial \over \partial {z} } \psi_2(x,y).
\end{eqnarray*}
Up to a scaling of the eigenvalue $\lambda$, these coupled equations are
themselves equivalent to the two-dimensional free particle Dirac equation.
Indeed, such a relationship between the two-dimensional Coulomb gas at
$\Gamma = 2$ and the two-dimensional Dirac equation is well known
\cite{CJ87}; the special feature here is that the components of the wave functions
$\psi_1$ and $\psi_2$ must be doubly periodic. 

We now seek the explicit form of the eigenvalues. For this purpose we compute
the Fourier expansion of
\begin{equation}\label{13.gxy1}
g(x,y;q) := {\theta_4(\pi (x + i y)/L ;q) \over
\theta_1(\pi (x + i y)/L;q)}
\end{equation}
by determining the coefficients $g_n(y;q)$ such that
$$
g(x,y;q) = \sum_{n = - \infty}^\infty g_n(y;q) e^{\pi i (2n+1) x/L}.
$$
In fact these are known from our earlier study \cite{Fo90d}, where with
$ q = e^{- \pi W / L}$ it was shown
\begin{equation}\label{13.13}
g_n(y;q) = 2i {\theta_4(0;q) \over \theta_1'(0;q)}
{e^{-\pi (2n+1) y / L} \over 1 - q^{-(2n+1)} }
\left \{ \begin{array}{ll} 1, & - W < y < 0 \\
q^{-(2n+1)} , & 0 < y < W. \end{array} \right.
\end{equation}
From this the eigenvalues in (\ref{13.epp}) can be determined.

\begin{prop}
Let $\mu := \pi (2n + 1)/L$ and $v := 2 \pi / \lambda$. Then
for each $n \in \mathbf Z$ the eigenvalues are
determined by the roots of the equation
\begin{equation}\label{13.ll}
\cosh \Big ( W(\mu^2 + v^2)^{1/2} \Big ) - 1 = 0.
\end{equation}
\end{prop}

\noindent
Proof. \qquad 
We seek eigenfunctions of the form
$$
\psi_1(x,y) = a_n(y) e^{\pi i (2n + 1)x /L}, \qquad
\psi_2(x,y) = b_n(y) e^{\pi i (2n + 1)x/L}.
$$
Substituting these forms and (\ref{13.gxy1}), and
using the orthogonality of $\{ e^{\pi i (2n + 1)x/L} \}$ shows that
(\ref{13.epp}) reduces to
\begin{eqnarray*}
\lambda a_n(y') & = & {\pi \theta_1'(0;q) \over \theta_4(0;q)}
\int_0^Wdy \, b_n(y) g_n(y' - y;q) \\
\lambda b_n(y') & = & {\pi \theta_1'(0;q) \over \theta_4(0;q)}
\int_0^Wdy \, a_n(y) g_n(y - y';q).
\end{eqnarray*}
Substituting the explicit form (\ref{13.13}) for $g_n$ and differentiating
shows that
\begin{eqnarray}\label{13.dy}
\lambda {d \over dy} \Big ( a_n(y) e^{\pi (2n + 1)y/L} \Big ) & = &
- 2 \pi i b_n(y) e^{\pi (2n + 1)y/L} \nonumber \\
\lambda {d \over dy} \Big ( b_n(y) e^{- \pi (2n + 1)y/L} \Big ) & = &
 2 \pi i a_n(y) e^{- \pi (2n + 1)y/L}
\end{eqnarray}
which are to be solved subject to the conditions
\begin{equation}\label{13.cond}
{a_n(W) \over a_n(0)} = 1, \qquad
{b_n(W) \over b_n(0)} = 1.
\end{equation}
Solving the equations in (\ref{13.dy}) in terms of a linear combination of
exponential functions, and determining the unspecified constants according
to (\ref{13.cond}), gives the condition (\ref{13.ll}). \hfill $\square$

\medskip

Substituting the eigenvalues as specified by (\ref{13.ll}) in (\ref{13.xfs})
we see that the product over the roots can be carried out according to the
following general formula \cite{Fo92f,JT96}.

\begin{prop}\label{p13.12}
For $f(z)$ an analytic function of $z$ with zeros at $z = \gamma_j$,
$j \in \mathbf Z$, and a product expansion of the form
$$
f(z) = A \prod_j \Big ( 1 - {z \over \gamma_j} \Big )
$$
we have
$$
\sum_j  \log \Big ( 1 + {c \over \gamma_j} \Big ) = \log {f(-c) \over f(0)}.
$$
\end{prop}

Thus
\begin{eqnarray*}
\Xi_2(\zeta) & = & \Big ( \theta_4(0;q) \Big )^2
\prod_{n=1}^\infty \Big ( {\cosh (W ( \mu^2 + (2 \pi \zeta)^2 )^{1/2}) - 1
\over \cosh (W \mu) - 1} \Big )^2 \\
&& = \Big ( \theta_4(0;q) \Big )^2
\prod_{n=1}^\infty e^{2W( \mu^2 + (2 \pi \zeta)^2 )^{1/2} - 2W \mu}
\Big ( {1 - e^{-W( \mu^2 + (2 \pi \zeta)^2 )^{1/2}}
\over 1 - e^{-W \mu}} \Big )^4.
\end{eqnarray*}
We know from a previous study \cite{Fo92f} that
\begin{equation}\label{hmm}
4 \pi \sum_{n=1}^\infty \Big ( (\zeta^2 + (n - 1/2)^2/L^2 )^{1/2} -
(n - 1/2)/L \Big ) \: 
\mathop{\sim}\limits_{L \to \infty} \:
- \beta P L + {\pi \over 6 L}
\end{equation}
where $P$ denotes the renormalized pressure (a renormalization, by way of a
cutoff in the sum over $n$ in (\ref{hmm}) is needed because of the short distance
singularity of the Boltzmann factor for a positive and negative charge at
$\Gamma = 2$). Thus for $W,L \to \infty$
$$
\prod_{p=1}^\infty e^{2W( \mu^2 + (2 \pi \zeta)^2 )^{1/2} - 2W \mu}
\sim e^{LW \beta P} q^{1/6}, \qquad q:= e^{- \pi W / L}.
$$
Furthermore, since \cite{WW65}
$$
\theta_4(z;q) = \prod_{n=1}^\infty (1 - e^{2 i z} q^{2n - 1})
 (1 - e^{-2 i z} q^{2n - 1})(1 - q^{2n})
$$
we see that
$$
\Big (\theta_4(0;q) \Big )^2 \prod_{n=1}^\infty
\Big ( {1 - e^{-W( \mu^2 + (2 \pi \zeta)^2 )^{1/2}}
\over 1 - e^{-W \mu}} \Big )^4
\: \sim \:  \prod_{n=1}^\infty (1 - q^{2n})^2.
$$
Hence, for $W$ and $L$ large, 
\begin{equation}\label{13.dpas}
- \log \Xi_2(\zeta) \sim - LW \beta P +
2 \log \Big ( q^{1/12} \prod_{n=1}^\infty (1 - q^{2n}) \Big ).
\end{equation}
Notice that the finite size correction term is precisely the same
as exhibited by (\ref{11.fr2p}) for the corresponding one-component
plasma system.

\section{Universality}
\setcounter{equation}{0}
A continuum viewpoint of general two-dimensional Coulomb systems in their
conductive phase (such systems may include arbitrary short range non-Coulomb
potentials) reveals a relationship with Gaussian field theory
\cite{Fo91,JMP94,JT96}. The latter, in the case of doubly periodic boundary
conditions exhibits the constant term present in (\ref{11.fr2p}) and
(\ref{13.dpas}), and as such predicts that this term will be present in the
expansion of the free energy of a general conductive two-dimensional Coulomb
system in doubly periodic boundary conditions.

To expand on these points, let us begin by recalling some of the arguments
from \cite{JT96}. The underlying assumption is that the universal features of
the grand partition function $\Xi_C$ of a conductive Coulomb system are correctly
accounted for by the continuum functional expression
\begin{equation}\label{14.01}
\Xi_C = \int {\cal D} \rho \,
\exp \Big ( - {\beta \over 2} \int \int d\vec{r} d\vec{r}\,' \,
\rho (\vec{r}) G(\vec{r},\vec{r}\,') \rho(\vec{r}\,') \Big ).
\end{equation}
Here $\rho (\vec{r})$ is the continuum charge density and $G(\vec{r},
\vec{r}\,')$ is the 2d Coulomb potential, defined as the solution of the
Poisson equation (\ref{5.0}) subject to doubly periodic boundary
conditions. The measure  ${\cal D} \rho$ normalized so that for
$ G(\vec{r},\vec{r}\,')=1$, $\Xi_C = 1$.

The expression (\ref{14.01}) is the continuum analogue of 
a multidimensional Gaussian integral. As such, taking into consideration
the normalization of ${\cal D} \rho$, it has the evaluation
$$
\Xi_C = (\det [ G(\vec{r},\vec{r}\,')])^{-1/2}.
$$
But from the Poisson equation $ G(\vec{r},\vec{r}\,') = (-{1 \over 2 \pi}
\nabla^2)^{-1}$, so this can be written as
\begin{equation}\label{14.02}
\Xi_C = \Big ( \det  \Big (- {1 \over 2 \pi} \nabla^2 \Big ) \Big )^{1/2}.
\end{equation}

On the other hand, the expression $(-{1 \over 2 \pi}
\nabla^2)^{-1/2}$ occurs in Gaussian field theory.
The partition function for such a theory is defined by the functional
integral
\begin{equation}\label{14.03}
Z_G  =  \int {\cal D} \phi \, \exp \Big ( - {\beta \over 2 \pi}
\int \phi(\vec{r}) ( - \nabla^2 ) \phi (\vec{r}) \, d \vec{r} \Big ) .
\end{equation}
Here it is required that $\phi (\vec{r})$ has doubly periodic boundary
conditions so as to be consistent with the Coulomb system, and
the normalization of $ {\cal D} \phi$ is chosen such that if
$ - \nabla^2$ is replaced by unity, then $Z_G=1$. 
This is evaluated by
diagonalizing $ - \nabla^2$, which shows 
\begin{equation}\label{14.04}
Z_G =  \Big ( \det  \Big (- {1 \over 2 \pi} \nabla^2 \Big ) \Big )^{-1/2},
\end{equation}
Comparing (\ref{14.02}) and (\ref{14.04}) shows
\begin{equation}\label{XZ}
\log \Xi_C = - \log Z_G.
\end{equation}

For the two-dimensional Gaussian free field in doubly periodic boundary
conditions, $Z_G$ as specified by (\ref{14.04}) has been evaluated by
Cardy \cite{Ca89}. Ignoring the zero eigenvalue (this has its origin in the
need to regularise the Poisson equation (\ref{5.0}) according to
(\ref{6.2}) for a doubly periodic solution to exist) it was shown
\begin{equation}\label{XZ1}
\log Z_G = 2 \log \Big ( q^{1/12} \prod_{n=1}^\infty (1 - q^{2n}) \Big ).
\end{equation}
This is the $O(1)$ term in 
(\ref{11.fr2p}) and
(\ref{13.dpas}), as consistent with (\ref{XZ}), and so establishes the
stated universality property.

\section*{Acknowledgement}
I thank B.~Jancovici for pointing out (some time ago) the expression
(\ref{XZ1}) from the work of Cardy, and furthermore for detailed comments
on the manuscript.
This work was supported by the Australian Research Council.


\end{document}